\documentclass[twocolumn]{aastex631}

\shorttitle{Solar Measurement of the CHIME Beam}
\shortauthors{The CHIME Collaboration}
\graphicspath{{./}{figures/}}

\begin{document}

\title{Using the Sun to Measure the Primary Beam Response of the Canadian Hydrogen Intensity Mapping Experiment}

\shortauthors{The CHIME Collaboration}

\correspondingauthor{Dallas Wulf}
\email{ dallas.wulf@mail.mcgill.ca }

\newcommand{\UBC}{Department of Physics and Astronomy, University of British Columbia, Vancouver, BC, Canada}
\newcommand{\MITP} {Department of Physics, Massachusetts Institute of Technology, Cambridge, MA, USA}
\newcommand{\MITK} {MIT Kavli Institute for Astrophysics and Space Research, Massachusetts Institute of Technology, Cambridge, MA, USA}
\newcommand{\TRU}{Department of Physical Sciences, Thompson Rivers University, Kamloops, BC, Canada}
\newcommand{\PI}{Perimeter Institute for Theoretical Physics, Waterloo, ON, Canada}
\newcommand{\DRAO}{Dominion Radio Astrophysical Observatory, Herzberg Astronomy \& Astrophysics Research Centre, National Research Council Canada, Penticton, BC, Canada}
\newcommand{\UBCO}{Department of Computer Science, Math, Physics, and Statistics, University of British Columbia-Okanagan, Kelowna, BC, Canada}
\newcommand{\McGill}{Department of Physics, McGill University, Montreal, QC, Canada}
\newcommand{\UofTastro}{David A.\ Dunlap Department of Astronomy \& Astrophysics, University of Toronto, Toronto, ON, Canada}
\newcommand{\UofTphys}{Department of Physics, University of Toronto, Toronto, ON, Canada}
\newcommand{\WVU} {Department of Computer Science and Electrical Engineering, West Virginia University, Morgantown WV, USA}
\newcommand{\WVUA} {Department of Physics and Astronomy, West Virginia University, Morgantown, WV, USA}
\newcommand{\WVUGWAC} {Center for Gravitational Waves and Cosmology, West Virginia University, Morgantown, WV, USA}
\newcommand{\Yale}{Department of Physics, Yale University, New Haven, CT, USA}
\newcommand{\YaleA}{Department of Astronomy, Yale University, New Haven, CT, USA}
\newcommand{\Dunlap}{Dunlap Institute for Astronomy and Astrophysics, University of Toronto, Toronto, ON, Canada}
\newcommand{\RRI}{Raman Research Institute, Sadashivanagar,   Bengaluru, India}
\newcommand{\ASIAA}{Institute of Astronomy and Astrophysics, Academia Sinica, Taipei, Taiwan}
\newcommand{\CITA}{Canadian Institute for Theoretical Astrophysics, Toronto, ON, Canada}
\newcommand{\CIFAR}{Canadian Institute for Advanced Research, 180 Dundas St West, Toronto, ON, Canada }
\newcommand{\WVUphysastro} {Department of Physics and Astronomy, West Virginia University, Morgantown, WV, USA}

\collaboration{100}{The CHIME Collaboration:}
%
%

%
\author[0000-0001-6523-9029]{Mandana Amiri}
\affiliation{\UBC}
\author[0000-0003-3772-2798]{Kevin Bandura}
\affiliation{\WVU}
\author{Anja Boskovic}
\affiliation{\UBC}
\author[0000-0001-6509-8430]{Jean-François Cliche}
\affiliation{\McGill}
\author[0000-0001-8123-7322]{Meiling Deng}
\affiliation{\DRAO}
\affiliation{\PI}
\affiliation{\UBC}
\author[0000-0001-7166-6422]{Matt Dobbs}
\affiliation{\McGill}
\author[0000-0002-6899-1176]{Mateus Fandino}
\affiliation{\UBC}
\affiliation{\TRU}
\author[0000-0002-0190-2271]{Simon Foreman}
\affiliation{\PI}
\affiliation{\DRAO}
\author[0000-0002-1760-0868]{Mark Halpern}
\affiliation{\UBC}
\author[0000-0001-7301-5666]{Alex S. Hill}
\affiliation{\UBCO}
\affiliation{\DRAO}
\author[0000-0002-4241-8320]{Gary Hinshaw}
\affiliation{\UBC}
\author[0000-0003-4887-8114]{Carolin H\"ofer}
\affiliation{\UBC}
\author[0000-0002-3354-3859]{Joseph Kania}
\affiliation{\WVUphysastro}
\author[0000-0003-1455-2546]{T.L. Landecker}
\affiliation{\DRAO}
\author[0000-0001-8064-6116]{Joshua MacEachern}
\affiliation{\UBC}
\author[0000-0002-4279-6946]{Kiyoshi Masui}
\affiliation{\MITK}
\affiliation{\MITP}
\author[0000-0002-0772-9326]{Juan Mena-Parra}
\affiliation{\MITK}
\author[0000-0002-7333-5552]{Laura Newburgh}
\affiliation{\Yale}
\author[0000-0002-2465-8937]{Anna Ordog}
\affiliation{\UBCO}
\affiliation{\DRAO}
\author[0000-0002-9516-3245]{Tristan Pinsonneault-Marotte}
\affiliation{\UBC}
\author[0000-0002-5283-933X]{Ava Polzin}
\affiliation{\YaleA}
\author[0000-0001-6967-7253]{Alex Reda}
\affiliation{\Yale}
\author[0000-0002-4543-4588]{J. Richard Shaw}
\affiliation{\UBC}
\author[0000-0003-2631-6217]{Seth R. Siegel}
\affiliation{\McGill}
\author[0000-0001-7755-902X]{Saurabh Singh}
\affiliation{\McGill}
\affiliation{\RRI}
\author[0000-0003-4535-9378]{Keith Vanderlinde}
\affiliation{\UofTastro}
\affiliation{\Dunlap}
\author[0000-0002-1491-3738]{Haochen Wang}
\affiliation{\MITK}
\affiliation{\MITP}
\author{James S. Willis}
\affiliation{\Dunlap}
\author[0000-0001-7314-9496]{{\rm and}  Dallas Wulf}
\affiliation{\McGill}

%
%
%





\begin{abstract}
We present a beam pattern measurement of the Canadian Hydrogen Intensity Mapping Experiment (CHIME) made using the Sun as a calibration source. As CHIME is a pure drift scan instrument, we rely on the seasonal North-South motion of the Sun to probe the beam at different elevations.  This semiannual range in elevation, combined with the radio brightness of the Sun, enables a beam measurement which spans $\sim$7,200 square degrees on the sky without the need to move the telescope. We take advantage of observations made near solar minimum to minimize the impact of solar variability, which is observed to be $<$10\% in intensity over the observation period. The resulting data set is highly complementary to other CHIME beam measurements---both in terms of angular coverage and systematics---and plays an important role in the ongoing program to characterize the CHIME primary beam.
\end{abstract}



\section{Introduction} \label{sec:intro}

The Canadian Hydrogen Intensity Mapping Experiment (CHIME) is a drift scan radio interferometer array operating between 400 and 800~MHz. It consists of four parabolic cylindrical reflectors.  Each reflector is 20~m in diameter, 100~m in length, oriented North-South, and is instrumented with 256 dual polarization feeds that are sensitive to linear polarizations in the North-South and East-West directions (1024 total feeds). This configuration provides a large instantaneous field of view, with the main lobe spanning $\sim$300 square degrees along the local meridian. CHIME is thus optimized for high mapping speed, making it a powerful instrument for 21~cm cosmology, transient detection, and source monitoring.  For a more detailed description of CHIME, we refer the reader to \cite{the_chime_collaboration_overview_2022} and references therein.

One of the challenges of working with CHIME is characterizing the large primary beam without the ability to steer the telescope. Due to the fixed pointing of the telescope, astrophysical radio sources that are typically used for beam measurements only transit the beam at a single declination. Thus, each source probes only a small fraction of the total beam solid angle. Moreover, few of these sources are bright enough to rise above the confusion level and be seen in the far side lobes. These challenges are shared by all drift scan instruments with limited pointing capability, such as the Molonglo Observatory Synthesis Telescope \citep[MOST,][]{bailes_utmost_2017}, the Deep Synoptic Array \citep[DSA-110,][]{kocz_dsa-10_2019}, the Canadian Hydrogen Observatory and Radio transient Detector \citep[CHORD,][]{vanderlinde_canadian_2019}, the Tianlai Project \citep{phan_overview_2020}, and the Hydrogen Intensity and Real-time Analysis eXperiment \citep[HIRAX,][]{crichton_hydrogen_2022}.  While some of these instruments are exploring the use of drones to overcome these challenges \citep[E.g.][]{zhang_beam_2021}, they are not feasible for CHIME, which has a far-field distance greater than one kilometer.

These considerations motivated us to use the Sun as a calibration source for measuring the primary beam response of CHIME.  The Sun has been used for beam-mapping purposes before \citep[E.g.][]{pauliny-toth_survey_1962, murphy_beam_1993, higgs_low-resolution_2000, chang_beam_2015}.  However, these measurements were made with fully steerable dishes, capable of measuring the full beam in a single observation by employing a raster scan strategy.  In contrast, the measurement reported here relies on the seasonal motion of the Sun and combines observations spanning one year. Between the winter and summer solstices, the declination of the Sun moves between $-23.5^{\circ}$  and $+23.5^{\circ}$, enabling a quasi-continuous measurement of the beam over this range.  Moreover, the radio brightness of the Sun exceeds 200~kJy in our band, enabling a high signal-to-noise measurement of the beam into the far side lobes.  These combined properties enable the Sun to probe the beam over $\sim$7,200 square degrees on the sky without the need to move the telescope. 

In what follows, we describe a measurement of the CHIME primary beam made using the Sun.  In Sec.~\ref{sec:sun}, we review the emission characteristics of the Sun that are relevant to its use as a calibration source at decimeter wavelengths. In Section~\ref{sec:methods} we describe the observations and analysis techniques used to produce a beam measurement.  Results are presented in Sec.~\ref{sec:results}, followed by a discussion of this measurement in the context of the broader CHIME beam measurement program in Sec.~\ref{sec:discuss}.

\subsection{Solar Emission at Decimeter Wavelengths} \label{sec:sun}

Solar emission at radio wavelengths is conventionally divided into three components: a minimum/quiet component, a slowly varying component, and a burst component.  The relative contributions of the latter two fluctuate according to the well-known 11-year solar cycle.  The observations reported here were made in late 2019 and early 2020, near the minimum point between Solar Cycles 24 and 25\footnote{\raggedright{National Research Council Canada, \url{https://www.spaceweather.gc.ca/}}}. The quiet component is produced by thermal free-free (Bremsstrahlung) emission, which, at decimeter wavelengths, originates in the lower corona.  Due to steep temperature and density gradients in this part of the solar atmosphere, the brightness temperature of the Sun is a strong function of frequency, with higher frequencies probing lower altitudes and thus lower temperatures.

\begin{figure}
	\includegraphics[width=\columnwidth]{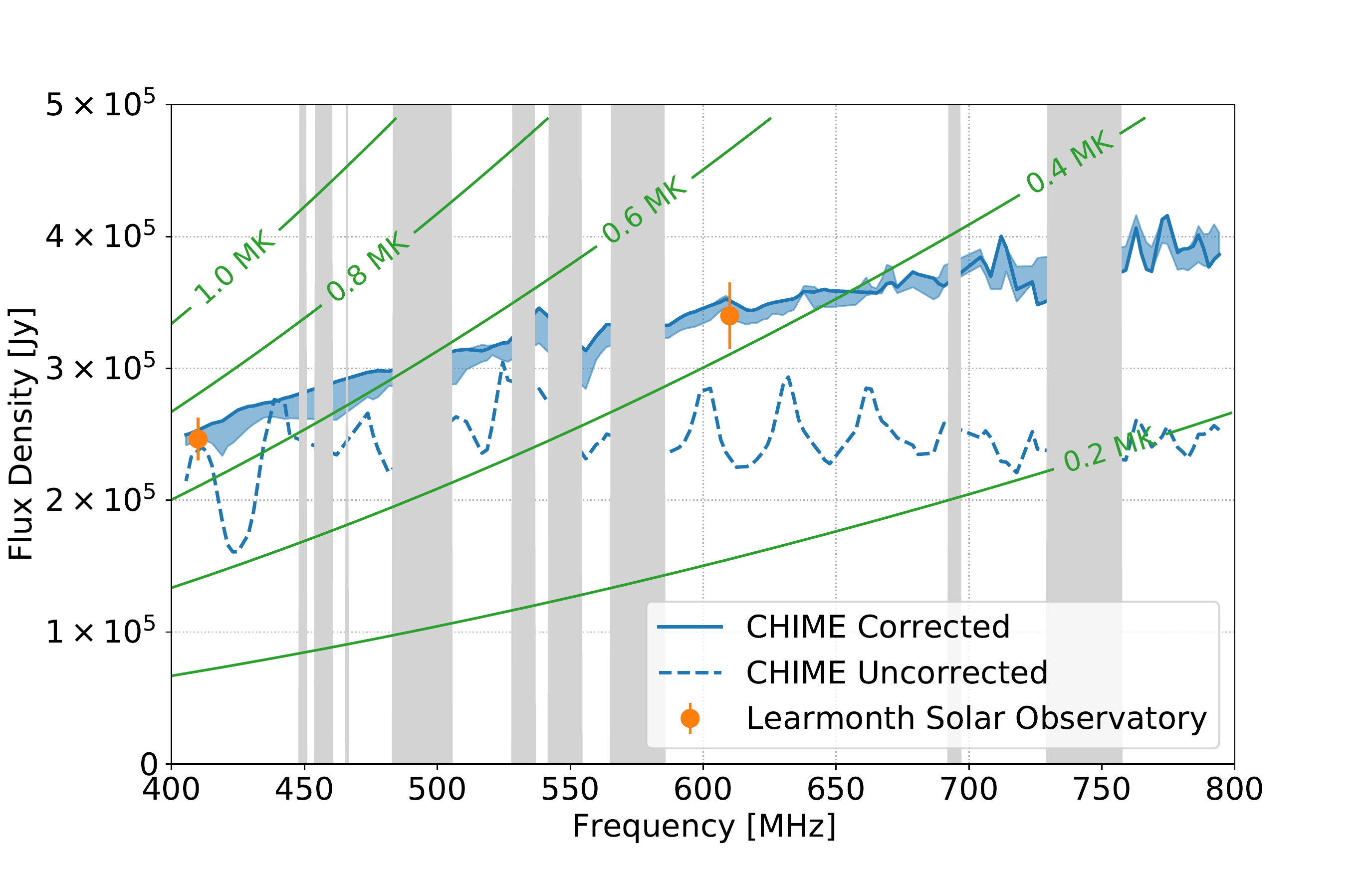}
    \caption{Total solar intensity calibrated against Tau~A on 31 May 2019, before (dashed blue) and after (solid blue) correcting for quantization bias (Section~\ref{sec:bias}). The shaded blue region covers the range of three other CHIME measurements made at the declinations of Tau~A and Vir~A through 11 July 2020.  Independent measurements of the mean and standard deviation of the daily mean solar flux are shown in orange \citep{SWS}. The gray bars mask frequencies of persistent RFI.  The green contours indicate the flux density of a 32 arcmin diameter disk with a given (frequency-independent) brightness temperature, and are shown to emphasize the frequency dependence of the effective temperature of the Sun in this band. Both CHIME and Learmonth observations indicate solar variability $<$10\% over the observation period.}
    \label{fig:spectrum}
\end{figure}

There is little direct imaging of the Sun in the CHIME band \citep{shibasaki_radio_2011}.  However, 432~MHz images from \citet{mercier_morphology_2012} made during the previous solar minimum (2008--2011) show that emission is nearly uniform over the face of the Sun and is confined to $\lesssim 1.1 R_s$, where \mbox{$R_s=16$} arcmin is the optical radius of the Sun as viewed from Earth.  At higher frequencies, the apparent radius is even closer to the optical radius. We therefore conclude that the emission profile of the Sun throughout the CHIME band can be approximated as a disk with radius $\sim R_s$. Given this radius and the temperature range of the lower corona, the flux density of the quiet Sun is a few $\times10^{5}$~Jy in the CHIME band, as shown in Figure~\ref{fig:spectrum}.

Even at solar minimum, there is still a contribution to the solar flux from slowly varying and burst components. To quantify the amplitude of these components, we refer to total intensity data from the Learmonth Solar Observatory \citep{SWS}, available at 1~second cadence at 410~MHz and 610~MHz. Between 31 May 2019 and 11 July 2020, the standard deviation of the daily mean flux at these frequencies was found to be 7\%  of the mean value, with comparable variation on intra-day timescales. Instrumental instability may account for a non-negligible fraction of the variability observed in the Learmonth data, though this contribution is likely small.  Nevertheless, we consider this to be an upper limit on the intrinsic solar variability. During this interval, CHIME made calibrated measurements of the solar spectrum four times on days when the Sun was at the declination of either Tau~A or Vir~A (31 May 2019, 20 August 2019, 21 April 2020, and 11 July 2020). The beam response at these declinations is independently measured using source fluxes from \cite{perley_accurate_2017}, enabling an absolute flux measurement of the Sun on these days. Averaged across the band, these four measurements have a range of 5\% relative to the mean.  Both the Learmonth and CHIME measurements are plotted in Figure~\ref{fig:spectrum}. More details about the CHIME flux measurements are described in Section~\ref{sec:flux}.

\section{Methods} \label{sec:methods}

\subsection{Data Set} \label{sec:data}

Observations of the Sun are extracted from the CHIME \texttt{stack} data set described in detail in \cite{the_chime_collaboration_overview_2022}.  CHIME is an interferometer which gathers data in the form of correlation products between pairs of feeds. We will therefore discuss the CHIME data in the language of interferometry, where a baseline is the separation between two feeds and a visibility is the correlation product of two voltage streams.  By design, the CHIME baseline configuration is highly redundant, enabling data compression by averaging over redundant baselines. Prior to averaging over baselines, data are averaged to 10~s cadence, calibrated for gain variations, and baseline pairs containing bad feeds are masked. Time and frequency bins impacted by radio frequency interference (RFI) are also masked. As these data are recorded as part of CHIME's primary science objectives, no dedicated observing time or specialized data products are needed for this measurement.

\subsection{Baseline Selection} \label{sec:baseline}

As described in Section~\ref{sec:sun}, the emission profile of the quiet Sun in the CHIME band is well described by a disk approximately 0.5$^{\circ}$ in diameter.  Due to its finite size, only baseline separations between 30~cm and 10~m are used to avoid resolving out the Sun. Zero-spaced baselines (autocorrelations) are masked to remove the receiver temperature bias.  Since the minimum East-West separation between feeds is 22~m (set by the spacing of the cylindrical reflectors), this baseline selection only includes purely North-South separations.  The corresponding synthesized beam has a North-South width of $\sim 2 ^{\circ}$ FWHM at 800~MHz and attenuates the solar flux by $<2\%$ at all frequencies. Finally, only co-polar visibilities are used, with North-South polarized data analyzed separately from East-West polarized data. After all baseline selection criteria have been applied, 32 unique baselines are used for each polarization.

\begin{figure}
	\includegraphics[width=\columnwidth]{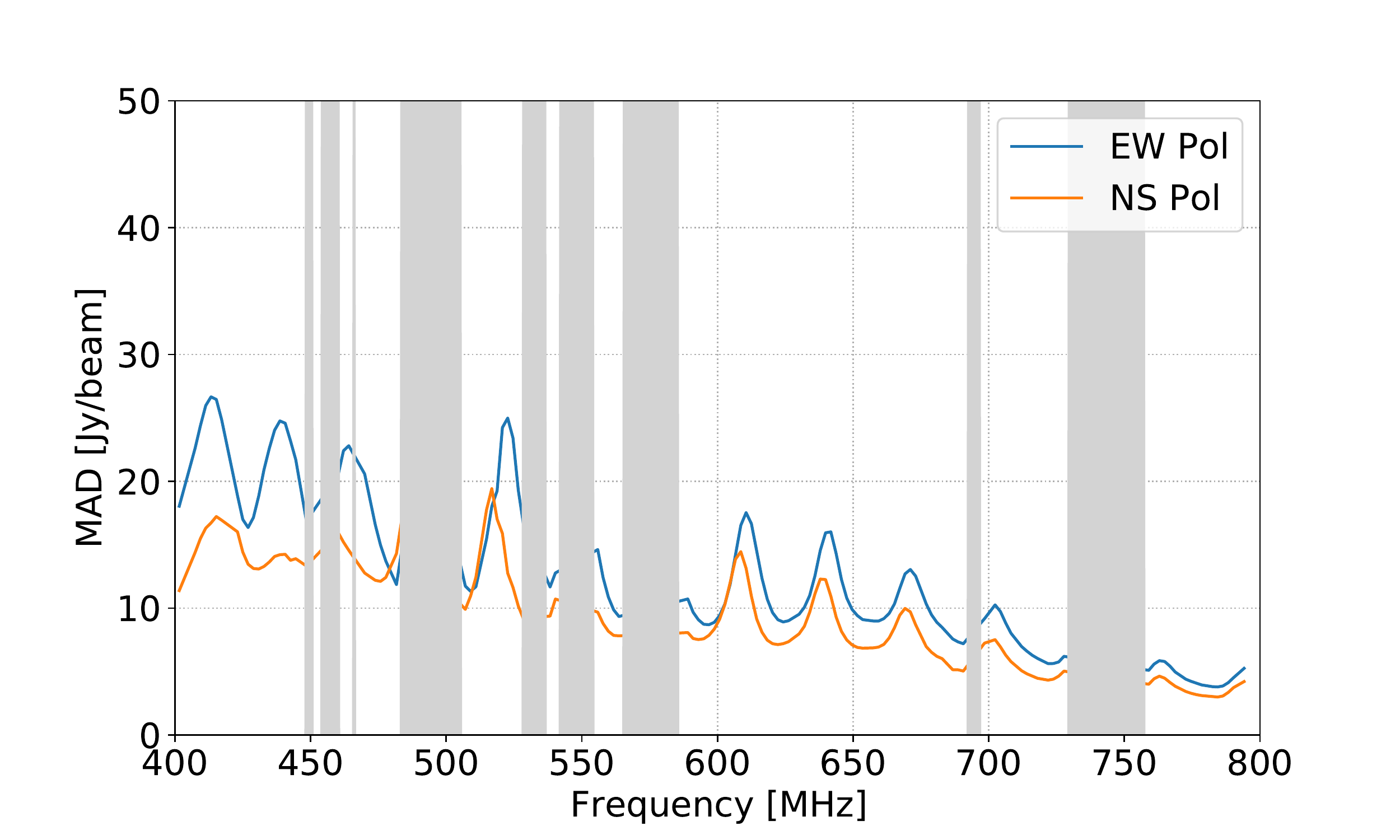}
    \caption{Estimate of confusion noise versus frequency for East-West polarized data (blue) and North-South polarized data (orange). The gray bars mask frequencies of persistent RFI. The difference between the two polarizations is primarily due to the different widths of the primary beams. Confusion noise is estimated as the median absolute deviation (MAD) of the night-time sky brightness over the solar declination range. On average, the confusion noise is about 4 orders of magnitude lower than the brightness of the Sun (Fig.~\ref{fig:spectrum}). There is a ripple with period 30~MHz due to standing waves between the reflector and the focal line \citep{the_chime_collaboration_overview_2022}.}
    \label{fig:confusion}
\end{figure}

The selection of baselines also affects the confusion noise limit.  Confusion noise is estimated by taking the median absolute deviation (MAD) of the night-time sky brightness over the solar elevation range, using a map made using the same baseline selection as the solar measurements (Figure~\ref{fig:confusion}).  Full sidereal coverage of the map is achieved by stitching together right ascension ranges observed over the course of the year.  Comparing Figures~\ref{fig:spectrum} and \ref{fig:confusion}, the flux of the Sun is seen to be about four orders of magnitude greater than the confusion noise across most of the band (ignoring beam effects, which are of order unity). Due to the geometry of the reflector, the East-West polarized beam is slightly wider than the North-South polarized beam, resulting in slightly higher confusion noise. The ratio of the solar flux to the confusion noise sets the dynamic range of the beam measurement, which is $\sim$40~dB.

\subsection{Beamforming} \label{sec:beamform}

The visibilities from selected baselines (after averaging over redundant baselines) are beamformed to the direction of the Sun.  Given the short baselines and large synthesized beam, we treat the Sun as a point source, beamforming only to its center position. The Sun is observed for the full duration that it is above the horizon on each available day between 31 May 2019 and 11 July 2020. Beamformed visibilities are then averaged over the selected baselines, weighted by the inverse noise variance of each visibility.  Visibilities in the \texttt{stack} data set have been calibrated in units of Jy/beam, such that this average also has units of Jy/beam. This interferometric view of the Sun results in a measurement of the average primary beam across the instrument.  Variations in the primary beam over individual feeds are not measured, as this would require the full set of redundant visibilities, which are not saved at all frequencies due to data volume considerations. 

\subsection{Flux Calibration} \label{sec:flux}

The flux of the Sun was calibrated once, when the Sun transited at the declination of Tau~A ($+22.7^{\circ}$) on 31 May 2019. At peak transit (0$^{\circ}$ hour angle) at this declination, the beam response for each polarization is independently measured using CHIME observations of Tau~A and the flux of Tau~A reported in \cite{perley_accurate_2017}.  These beam measurements allow us to calibrate the solar spectrum at this point in time. As discussed in Section~\ref{sec:sun}, the average solar flux in the CHIME band varied by $\lesssim$10\% between 31 May 2019 and 11 July 2020. Without simultaneous solar monitor data available across the band, we are unable to track the true flux of the Sun to better than this precision and so we treat the flux as constant in this analysis. Among the potential calibrators within the solar declination range, Tau~A is chosen for its high signal to noise.  Tau~A is the brightest radio source in this range and transits at an elevation where the CHIME primary beam response is also high. As a cross-check, the flux of the Sun was calibrated again against Tau~A on 11 July 2020 and against Virgo~A on 20 August 2019 and 21 April 2020. As shown in Figure~\ref{fig:spectrum}, these measurements agree to within 5\% across the band.

\subsection{Correcting for Quantization Bias} \label{sec:bias} 

The dynamic range of the CHIME correlator is limited by the bit depth of the channelized voltages, which are rounded to four real bits and four imaginary bits prior to correlation \citep{bandura_ice_2016}. While this bit depth is sufficient for CHIME's primary science objectives, it introduces a bias due to digital compression when the Sun is in the main lobe of the primary beam.  If left uncorrected, this effect can bias the solar spectrum by as much as $\sim$30\% in amplitude, as shown in Figure~\ref{fig:spectrum}.  We account for this effect using an approximation of the analytical corrections derived in \citet{benkevitch_van_2016} and \citet{mena-parra_quantization_2018} for complex digital correlators. For a complete description, the reader should refer to these publications, though we will briefly describe the process here. 

Equation~21 of \citet{benkevitch_van_2016} provides an analytical expression for the quantized correlation {$\hat{\kappa}_{ij}$} given the true, unquantized correlation, {$\kappa_{ij}$}, as well as the standard deviations of inputs $i$ and $j$. We will refer to this function as $g$.

\begin{equation}
    \hat{\kappa}_{ij} = g(\kappa_{ij},\sigma_{i},\sigma_{j})
    \label{eq:g}
\end{equation}

\noindent Similarly, Equation~8 of \citet{mena-parra_quantization_2018} provides an expression for the quantized standard deviation as a function, $f$, of the true standard deviation.

\begin{equation}
    \hat{\sigma} = f(\sigma)
    \label{eq:f}
\end{equation}

\noindent The objective is to recover the true correlation, $\kappa$, when only the quantized quanities, $\hat{\sigma}$ and $\hat{\kappa}$, are known.  This is accomplished by inverting Equations~\ref{eq:g} and \ref{eq:f} as follows.

\begin{eqnarray}
    \kappa_{ij} & = & g^{-1}(\hat{\kappa}_{ij},\sigma_{i},\sigma_{j}) \nonumber\\ & = & g^{-1}(\hat{\kappa}_{ij},f^{-1}(\hat{\sigma}_{i}),f^{-1}(\hat{\sigma}_{j}))
    \label{eq:invert}
\end{eqnarray}

Our analysis makes two further simplifying assumptions. The first assumption is to ignore the phase error of $\hat{\kappa}$ and only to correct for amplitude.  At peak signal level, the phase error in a single baseline due to compression is $\sim$10~arcmin and becomes negligible ($\lesssim$10~arcsec) when averaged over redundant baselines.  Second, because the CHIME \texttt{stack} data set has been averaged over redundant baselines, we need to assume that $\hat{\sigma}^{2}_{i}=\hat{\sigma}^{2}_{j}=\langle\hat{\sigma}^{2}\rangle$ and $\hat{\kappa}_{ij}=\langle\hat{\kappa}\rangle$ for all inputs $i,j$ in each unique baseline. However, since the CHIME off-line data set does include the full correlation matrix (which has not been averaged over redundant baselines) at four frequencies, we can compare the exact analytical correction to our simplified version at these frequencies. We find that the error incurred by averaging over redundant baselines is $\lesssim$1\%---less than the intrinsic flux variation of Sun.

\section{Results} \label{sec:results}

The power beam ratio is formed by dividing the measured response to the Sun in each direction by the calibrated flux of the Sun. We normalize the beam to unity at the transit of Cyg~A ($0^{\circ}$ hour angle, $+40.7^{\circ}$ declination) to be consistent with how the data are normalized by the complex gain calibration procedure (for details, see \citealt{the_chime_collaboration_overview_2022}). As a consequence, the beam will have values greater than one in some directions.

\begin{figure}
\includegraphics[width=\columnwidth]{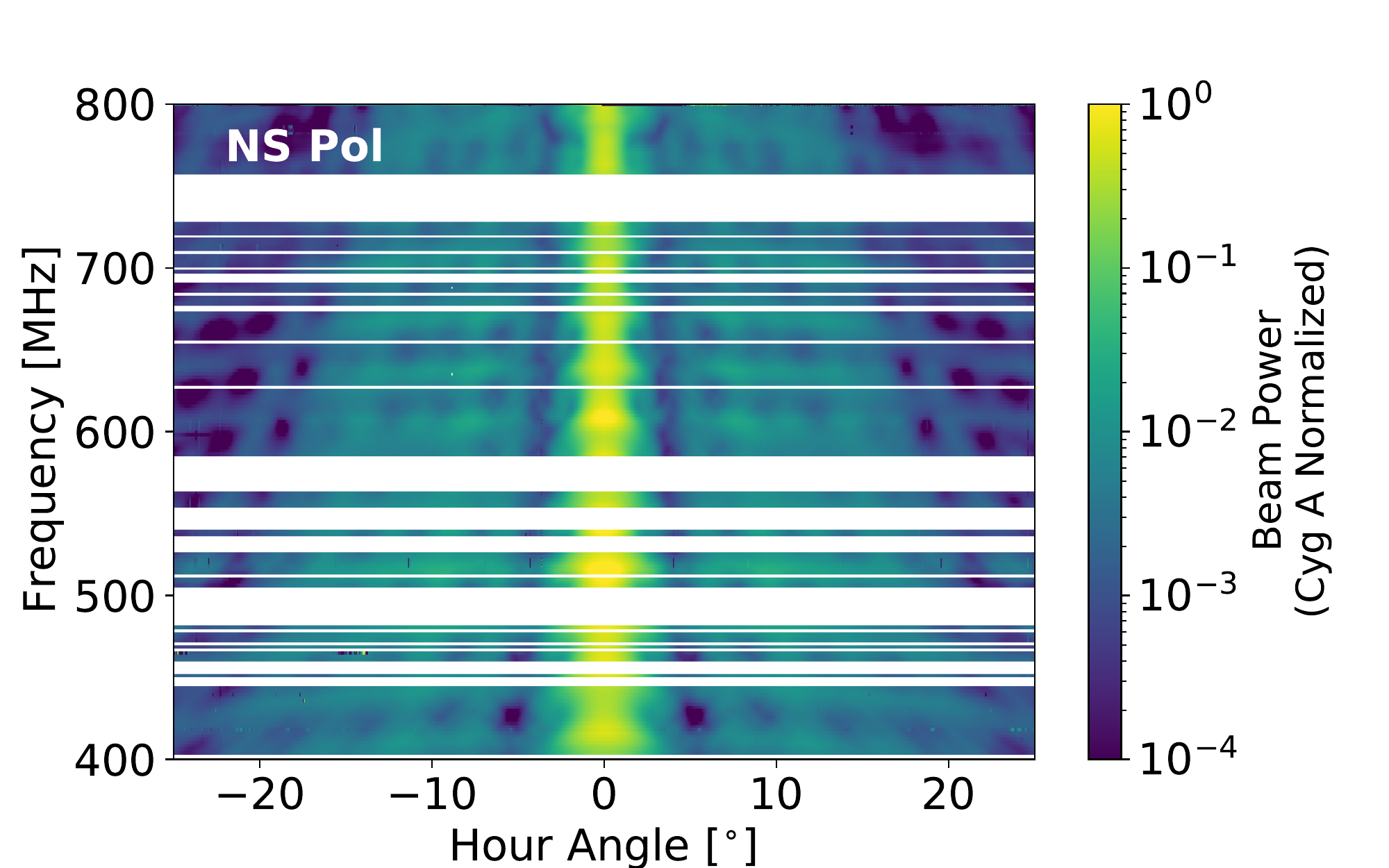}
\hspace{0mm}
\includegraphics[width=\columnwidth]{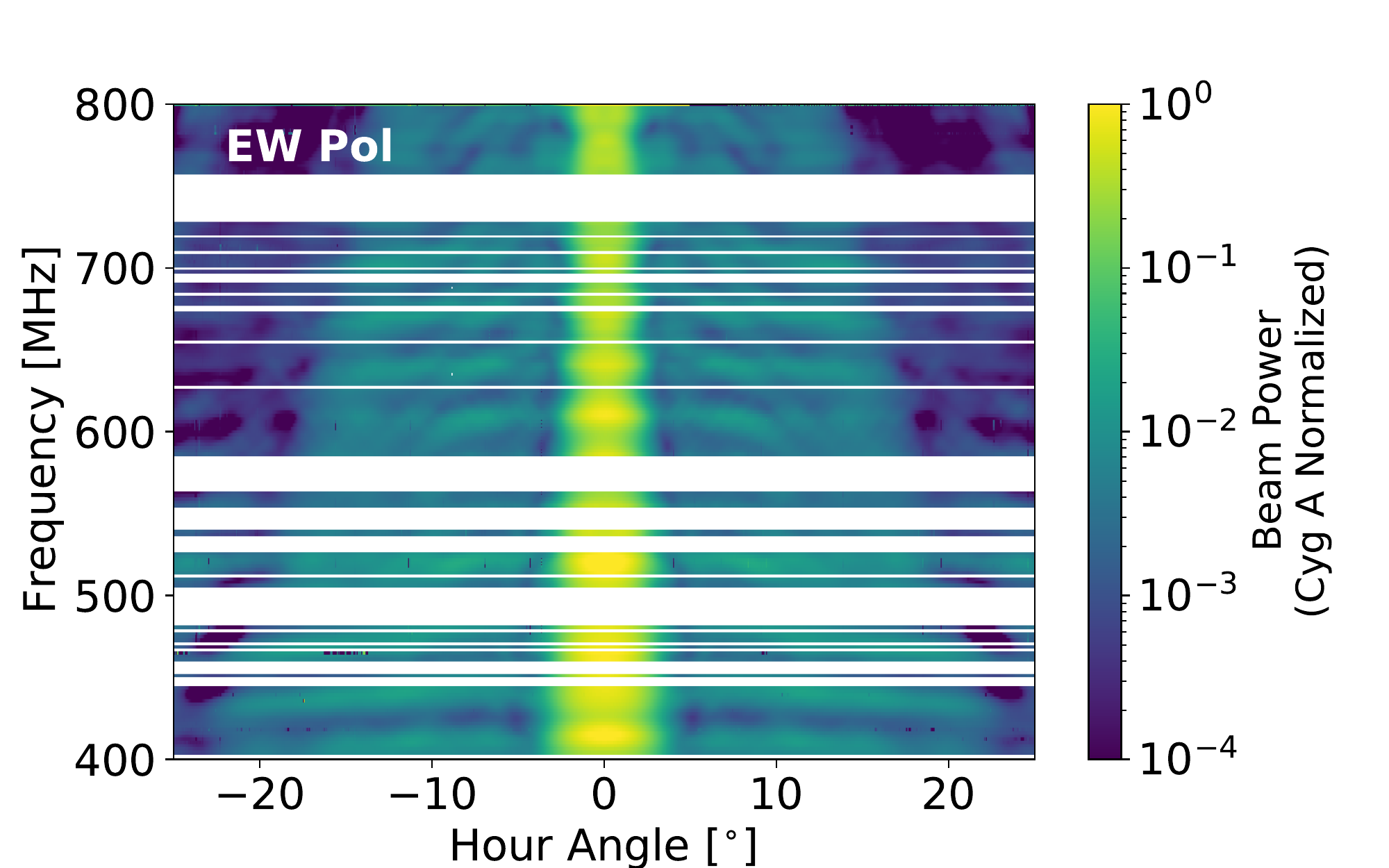}
\caption{North-South (top) and East-West (bottom) polarized beam measurements from a single solar transit on 20 August 2019, when the Sun was at the declination of Vir~A ($+12.4^{\circ}$). On large frequency scales, the beam width is proportional to wavelength, as expected due to diffraction. There is also a modulation in the beam amplitude and width on 30~MHz scales, due to standing waves between the reflector and the focal line. By convention in CHIME, the beam is normalized to be 1 at the transit of Cyg~A ($0^{\circ}$ hour angle, $+40.7^{\circ}$ declination)}
\label{fig:transit}
\end{figure}

Each solar transit traces a line of approximately constant declination. The transit on 20 August 2019, when the Sun was at the declination of Vir~A ($+12.4^{\circ}$), is shown in Figure~\ref{fig:transit}. Throughout the year, the daily change in solar declination is less than the $\sim$0.5$^{\circ}$~diameter of the Sun, enabling consecutive transits to be stitched together to achieve continuous declination coverage.  The full solar declination range was observed twice: first in late 2019 and again in early 2020.  Due to planned software deployments and telescope maintenance resulting in instrument downtime, certain declination ranges were missed in 2019.  In 2020, deployments were planned to avoid missing the same declination ranges twice, enabling complete declination coverage in the combined data set.  

\begin{figure}
\includegraphics[width=\columnwidth]{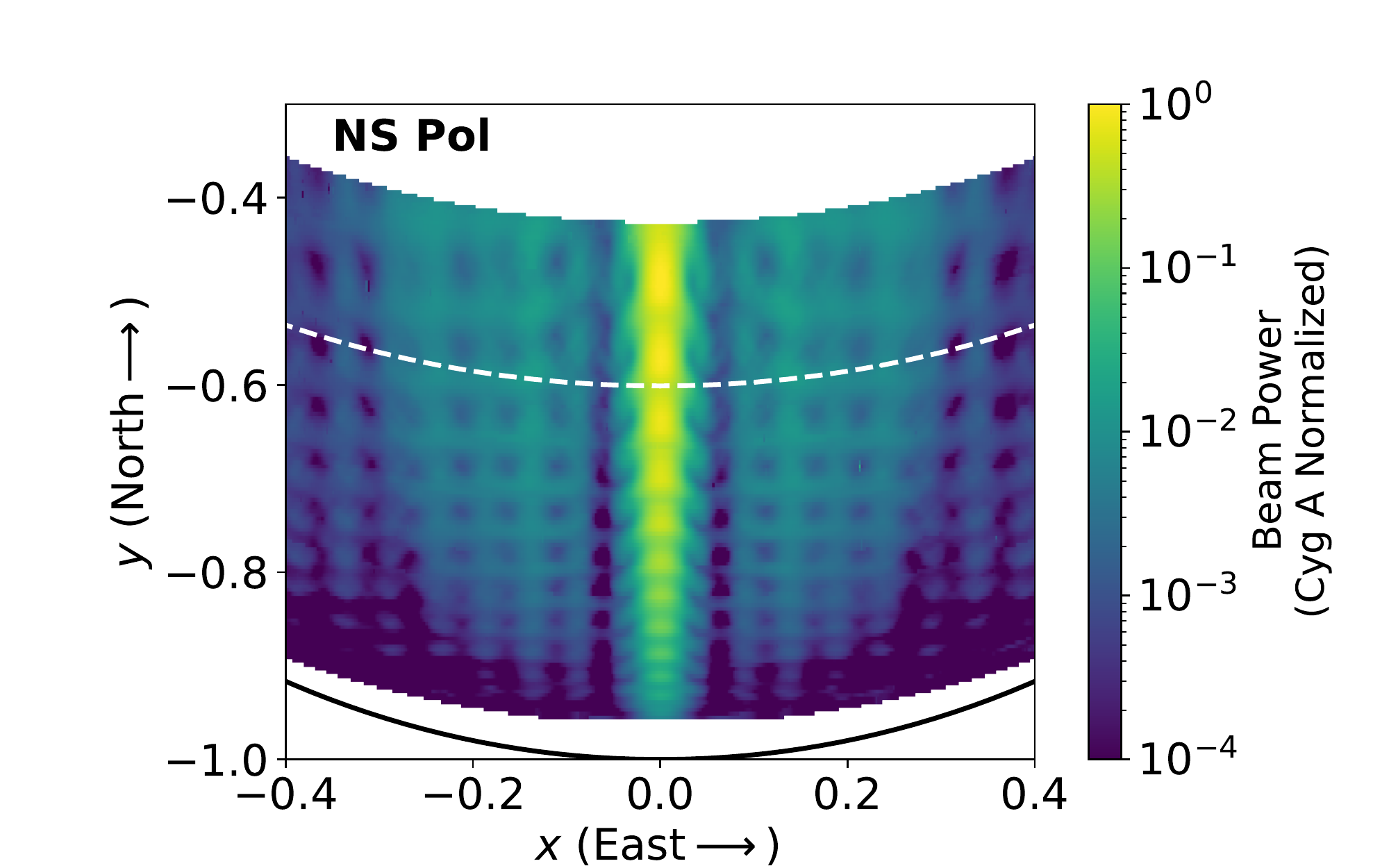}
\hspace{0mm}
\includegraphics[width=\columnwidth]{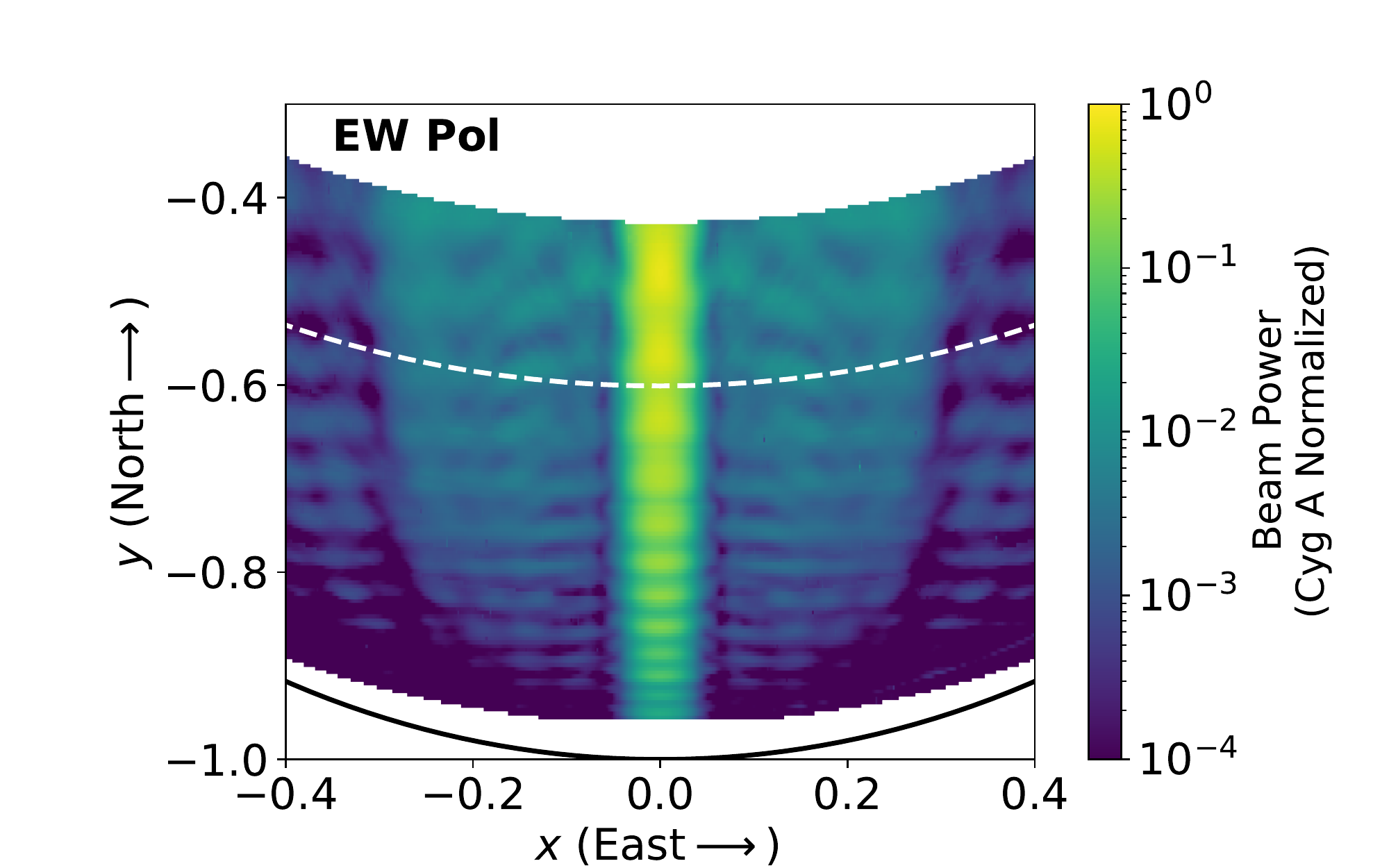}
\caption{Orthographic projection of the North-South (top) and East-West (bottom) polarized beam measurements at 600~MHz, made by combining one year of solar transits.  The modulation in beam width on small angular scales is real and arises from multiple reflection paths within the telescope. The white dashed line marks the path of the Sun made during the transit shown in Figure~\ref{fig:transit}.  The black curved line near the bottom of the plot marks the Southern horizon. The zenith (off scale) is located at the coordinate (0,0).}
\label{fig:beam}
\end{figure}

\begin{figure}
	\includegraphics[width=\columnwidth]{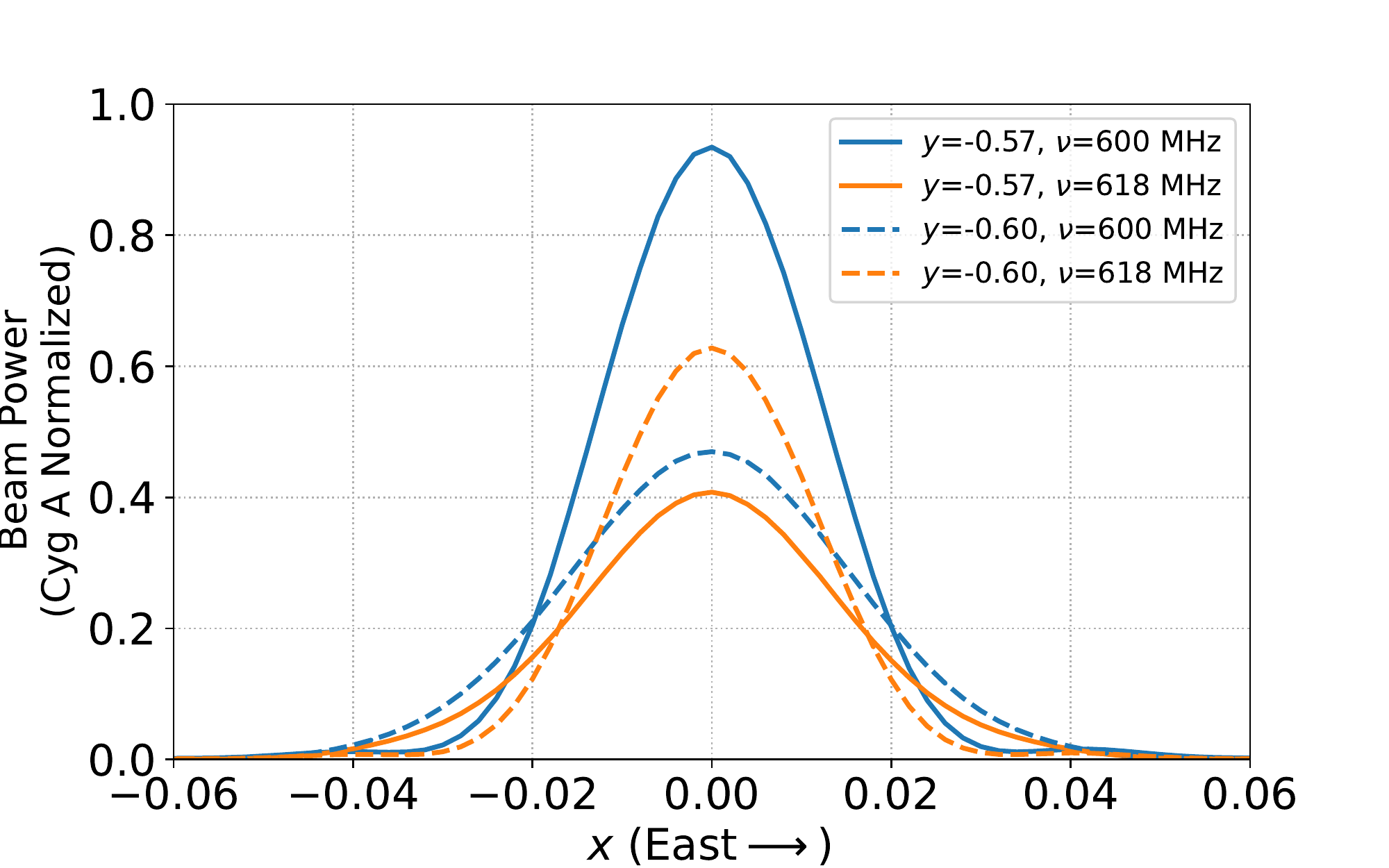}
    \caption{Slices of the North-South polarized beam as a function of orthographic coordinate $x$, emphasizing the evolution of the beam profile over frequency and elevation angle. Beam slices at 600~MHz (618~MHz) are plotted in blue (orange). Beam slices at $y=-0.57$ ($-0.60$) are plotted with solid (dashed) lines. The variations in beam amplitude and width arise due to interference between multiple reflection paths within the telescope. Rays with multiple reflections have broader illumination compared to the primary ray, resulting in narrower beam widths at frequencies and angles corresponding to constructive interference (solid blue and dashed orange lines).}
    \label{fig:beamwidths}
\end{figure}

The combined data set at 600~MHz is shown in Figure~\ref{fig:beam} using an orthographic projection with its origin at zenith. This projection has the advantage of not distorting the apparent beam width at different elevations and the projected coordinates $x$ and $y$ are always parallel to East and North, respectively. The coordinates $x$ and $y$ are the projections of the unit vector pointing to hour angle $h$ and declination $\delta$, given by

\begin{equation}
    x = - \cos{\delta} \sin{h} 
\end{equation}

and 

\begin{equation}
    y = \cos{\delta_0} \sin{\delta} - \sin{\delta_0} \cos{\delta} \cos{h}
\end{equation}

\noindent where $\delta_0$ is the latitude of the observer ($+49.3^{\circ}$ for CHIME). The basic shape of the beam in Figure~\ref{fig:beam} can be described as a 1$^{\circ}$-- 4$^{\circ}$ wide stripe along the local meridian, with moderate ($-20$~dB) side lobes extending up to $20^{\circ}$ on either side. As shown in Figure~\ref{fig:beamwidths}, the amplitude and width of the main lobe varies considerably ($\sim$25\%) over modest changes in frequency and elevation angle.  There is also significant structure in the side lobes of both polarizations.  Though difficult to quantitatively predict, these variations are qualitatively expected due to multiple reflection paths within the telescope, cross talk between feeds, and blockages in the aperture. For more in-depth discussion of these beam features, we refer the reader to \cite{the_chime_collaboration_overview_2022}.

\begin{figure}
\includegraphics[width=\columnwidth]{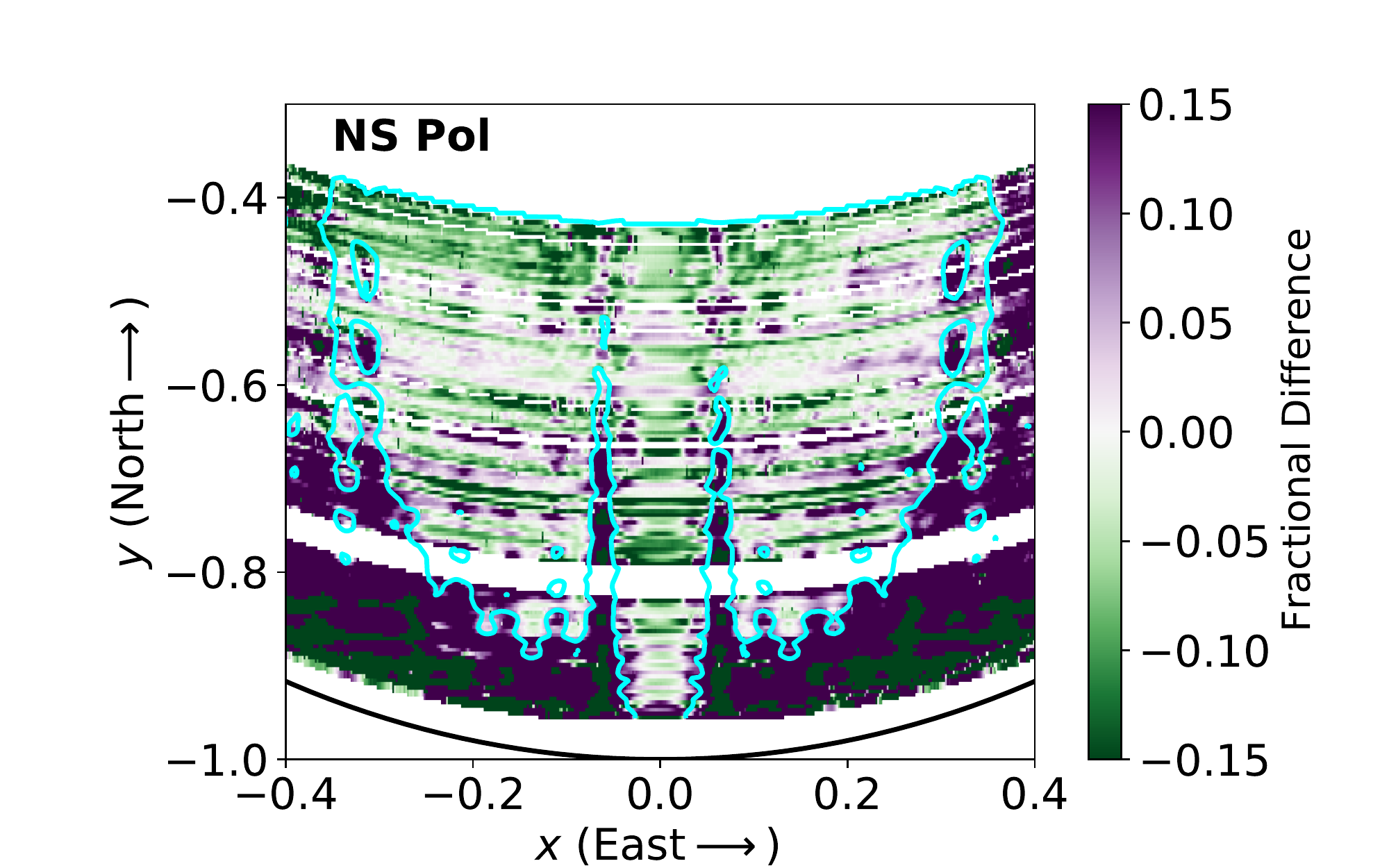}
\hspace{0mm}
\includegraphics[width=\columnwidth]{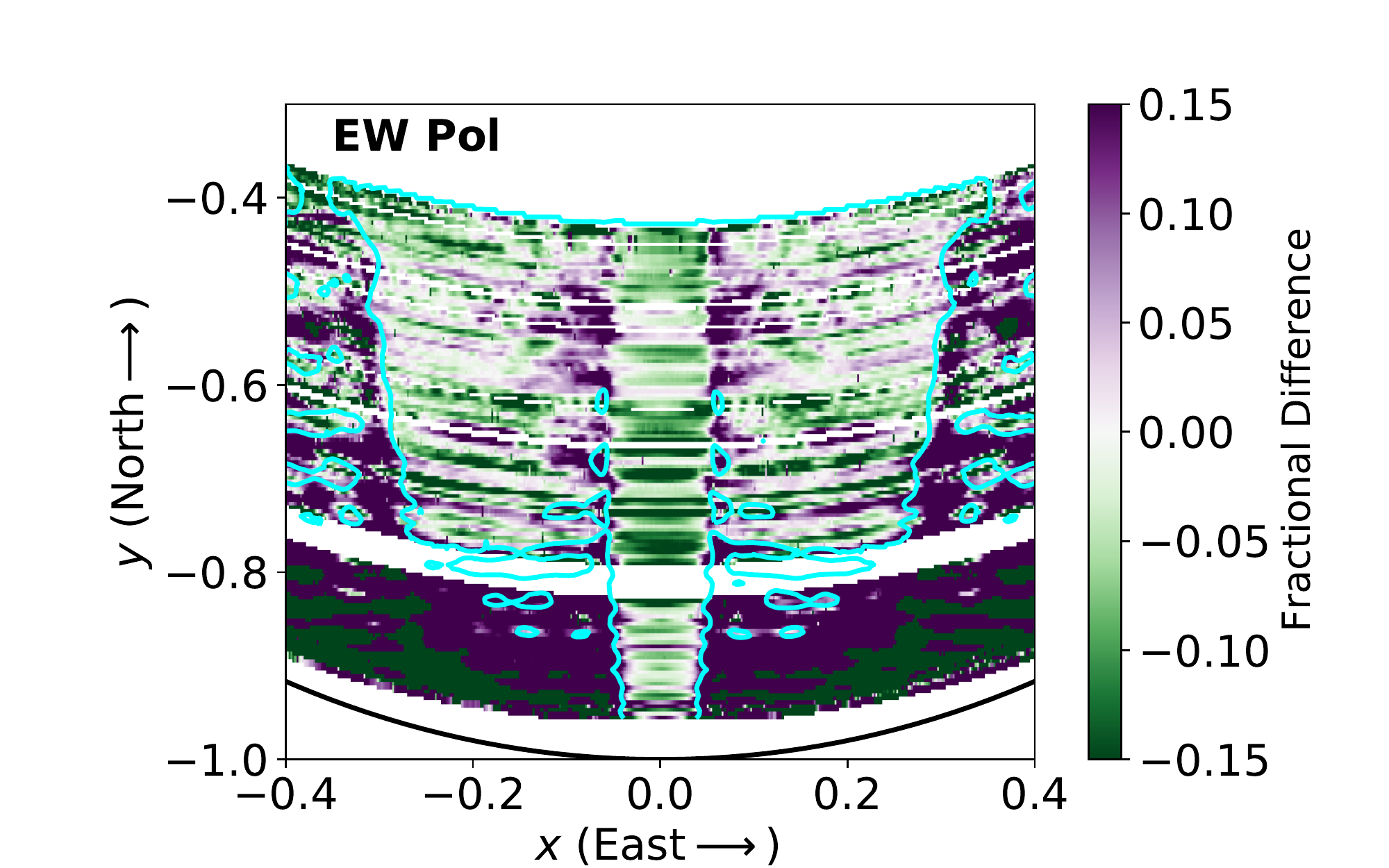}
\caption{The fractional difference between measurements made in late 2019 and early 2020, shown for the North-South (top) and East-West (bottom) polarized beam at 600~MHz.  White space corresponds to where only a single measurement exists. Where the beam response in Figure~\ref{fig:beam} is greater than $10^{-3}$ (cyan contour), the measurement uncertainty is dominated by intrinsic solar variability.  Below this level, confusion noise is comparable to or larger than the effect of solar variability. For both polarizations, the fractional differences within the contour approximately follow a normal distribution with $\sigma=9\%$.}
\label{fig:difference}
\end{figure}

Where repeat observations are available, the two measurements are averaged and their difference is used to assess measurement repeatability (Figure~\ref{fig:difference}). In regions where the response to the Sun is much greater than the confusion noise, we expect the difference between measurements to be dominated by variations in the flux of the Sun.  Comparing Figures~\ref{fig:beam} and \ref{fig:difference}, where the beam response is greater than $10^{-3}$ ($\gtrsim$10$\times$ the confusion limit), the fractional difference is found to be 9\% at 600 MHz, which is comparable with our estimate of solar variability over the observation period. Where the beam response is less than $10^{-3}$, the measurement becomes increasingly confusion-noise limited, causing the fractional error to increase.

\subsection{Comparison With Other Beam Measurements} \label{sec:compare}

The solar beam measurements are only part of the broader CHIME beam measurement program.  Although the Sun probes about 7,200 square degrees, additional measurements are needed to span the full sky.  Where spatial overlap exists between measurements, we can make comparisons. The other measurements all rely on astrophysical point source observations and thus have complementary systematics to the solar measurement. In general, the fluxes of these sources are more stable than the Sun (though they may have large uncertainties) and are small enough to not require correction for quantization bias.  Additionally, point source observations can be made with a complementary set of baselines, since these sources are unresolved by CHIME's longest baselines.

\begin{figure}
    \centering
    \includegraphics[width=\columnwidth]{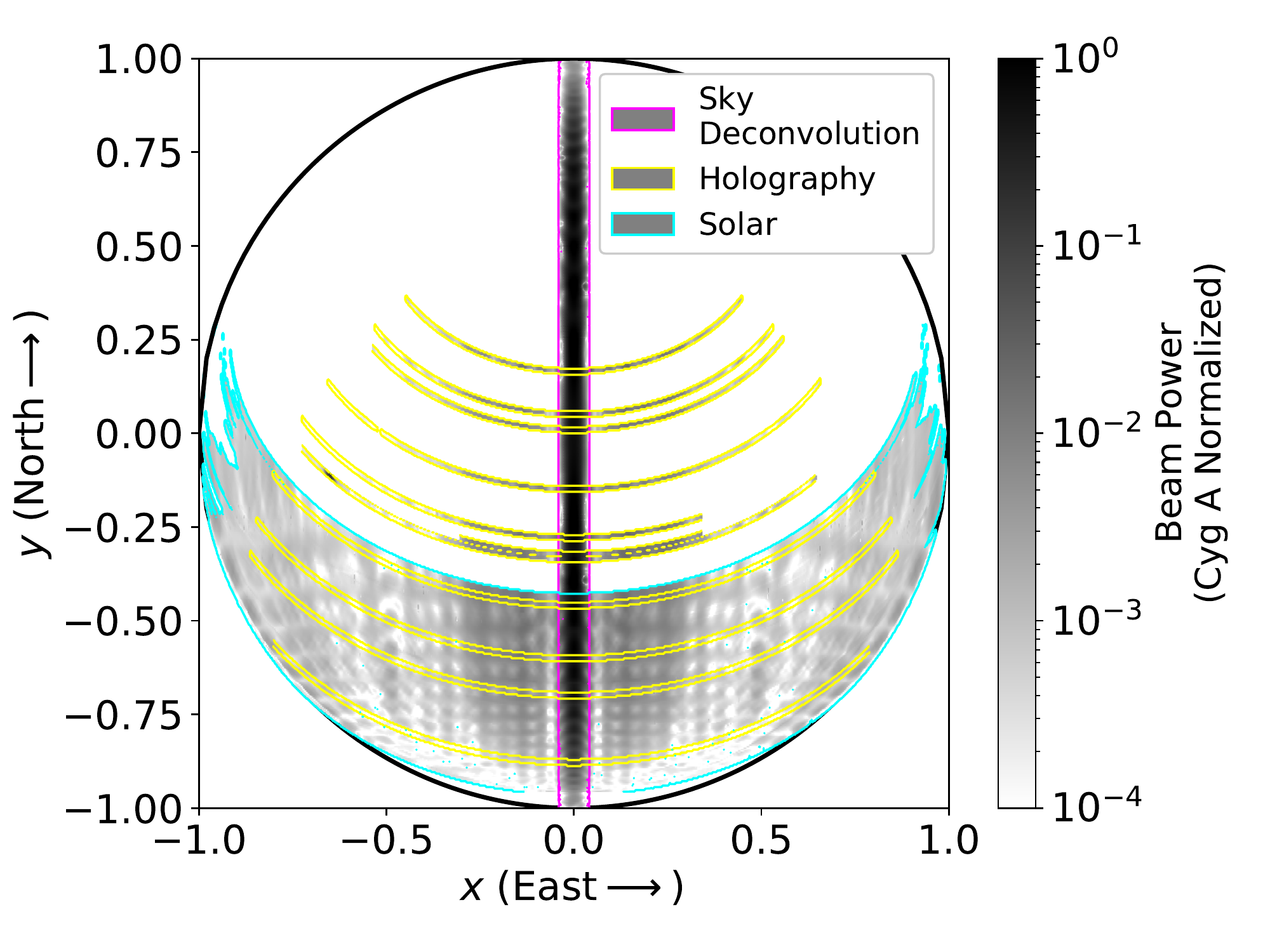}
    \caption{Orthographic projection of multiple North-South polarized beam measurements at 600~MHz. Despite spanning more than 7,200 square degrees, the solar measurement (cyan outline) only probes the main lobe between \mbox{$y=-0.44$} and \mbox{$y=-0.95$}.  In contrast, the beam measurement derived by deconvolving a point source model of the sky (magenta outline) spans the full elevation range of the main lobe, but does not probe the side lobes. Holography tracks of the brightest point sources (yellow outline) probe the beam at discrete elevations over a wide range in hour angle. The black circle marks the horizon.}
    \label{fig:allmeas}
\end{figure}

The angular coverage of the different beam measurements is summarized in Figure~\ref{fig:allmeas}.  There are two categories of measurements which rely on astrophysical point sources: source tracks and deconvolutions.  The former are made by beamforming visibilities to the location of bright sources as they transit through the primary beam, analogously to how the solar measurement is made. Beamforming may be done with CHIME-only baselines, or with baselines between CHIME and the John A. Galt 26~m tracking telescope \citep{locke_1420_1965} using the holographic technique described in \cite{berger_holographic_2016}. Holographic measurements have the added benefit of measuring the beam phase in addition to amplitude.  The second type of point source beam measurement is made by deconvolving a model of the point source sky from a CHIME map as described in \cite{stacking_2022}. Unlike the source tracks, which probe discrete declinations over a moderately wide range in hour angle, the deconvolved beam has continuous declination coverage, but is limited to the width of the main lobe in the East-West direction.  

\begin{figure}
\includegraphics[width=\columnwidth]{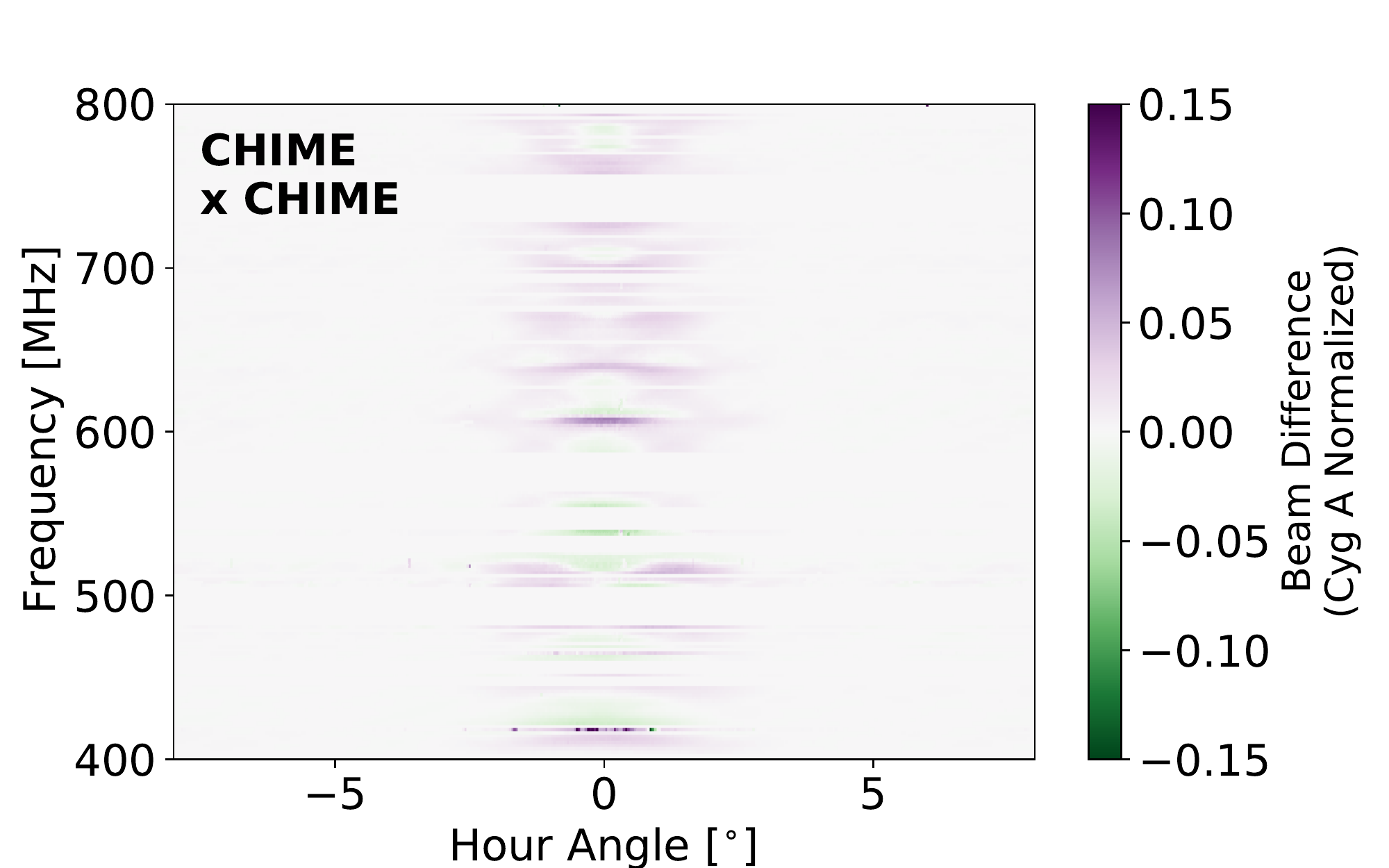}
\hspace{0mm}
\includegraphics[width=\columnwidth]{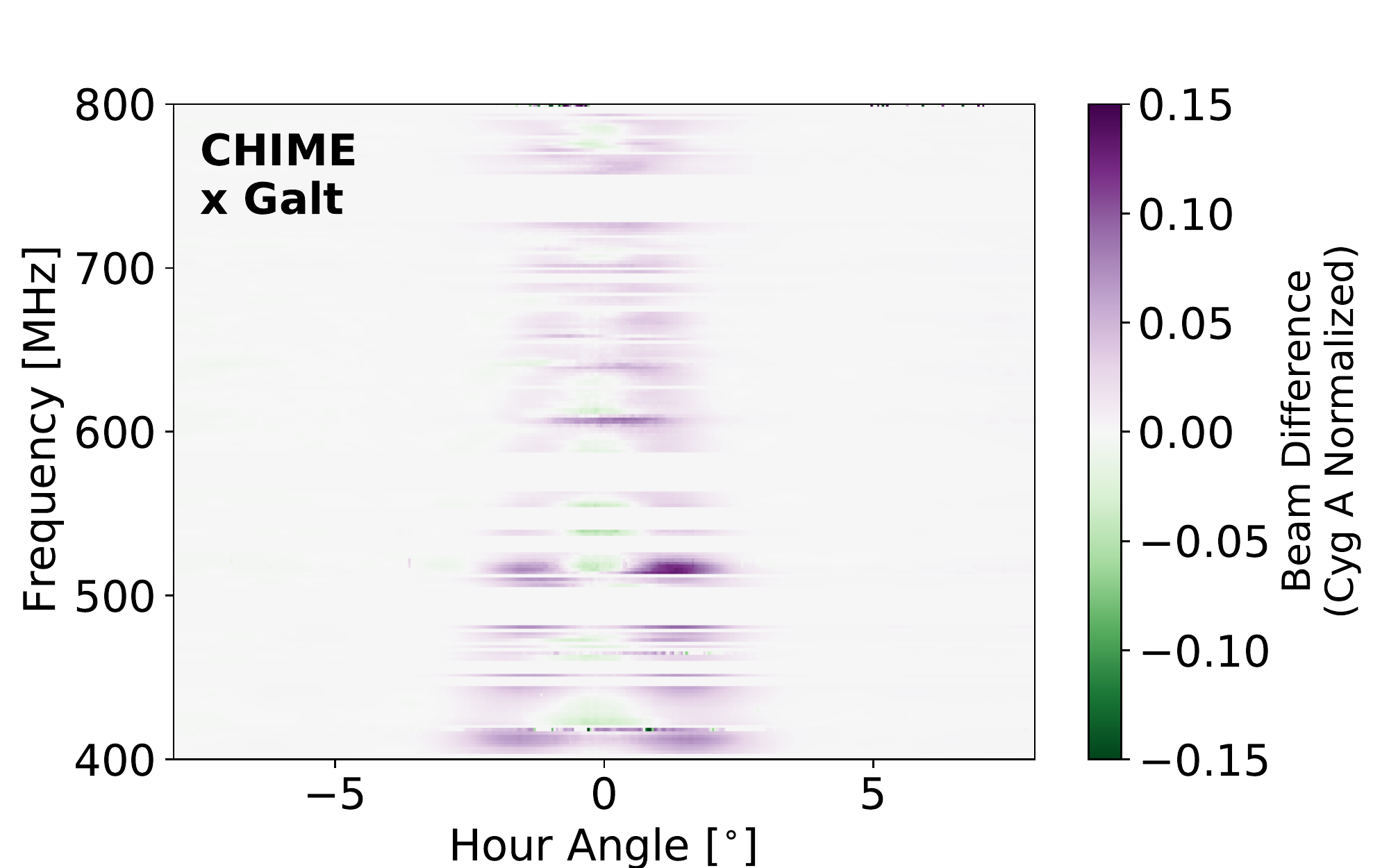}
\caption{The difference between the solar measurement of the North-South polarized beam at the declination of Vir~A ($+12.4^{\circ}$) and a CHIME-only measurement of Vir~A (top), and a holographic measurement of Vir~A (bottom). Systematic differences between beam measurements are typically less than a few percent of the peak beam response, and are primarily due to percent-level differences in measured beam width.  Systematic differences between the East-West polarized beam measurements (not shown) are approximately a factor of two larger.}
\label{fig:virA}
\end{figure}

In Figure~\ref{fig:virA}, we show the difference between the solar measurement at the declination of Vir~A ($+12.4^{\circ}$) and source tracks of Vir~A.  Note that, unlike Tau~A, Vir~A was not explicitly used to calibrate the flux of the Sun and thus provides a truly independent measurement of the beam.  While the CHIME-only observation of Vir~A provides slightly better agreement with the solar measurement than does the holographic measurement, both comparisons reveal systematic differences which are small---typically a few percent of the peak beam response. Residuals are dominated by systematic differences in the East-West width of main lobe. For the North-South polarized beams shown in Figure~\ref{fig:virA}, the solar beam is on average 2\% wider than the CHIME-only measurement of Vir-A, and 6\% wider than the holographic measurement. 

\begin{figure}
\includegraphics[width=0.9\columnwidth]{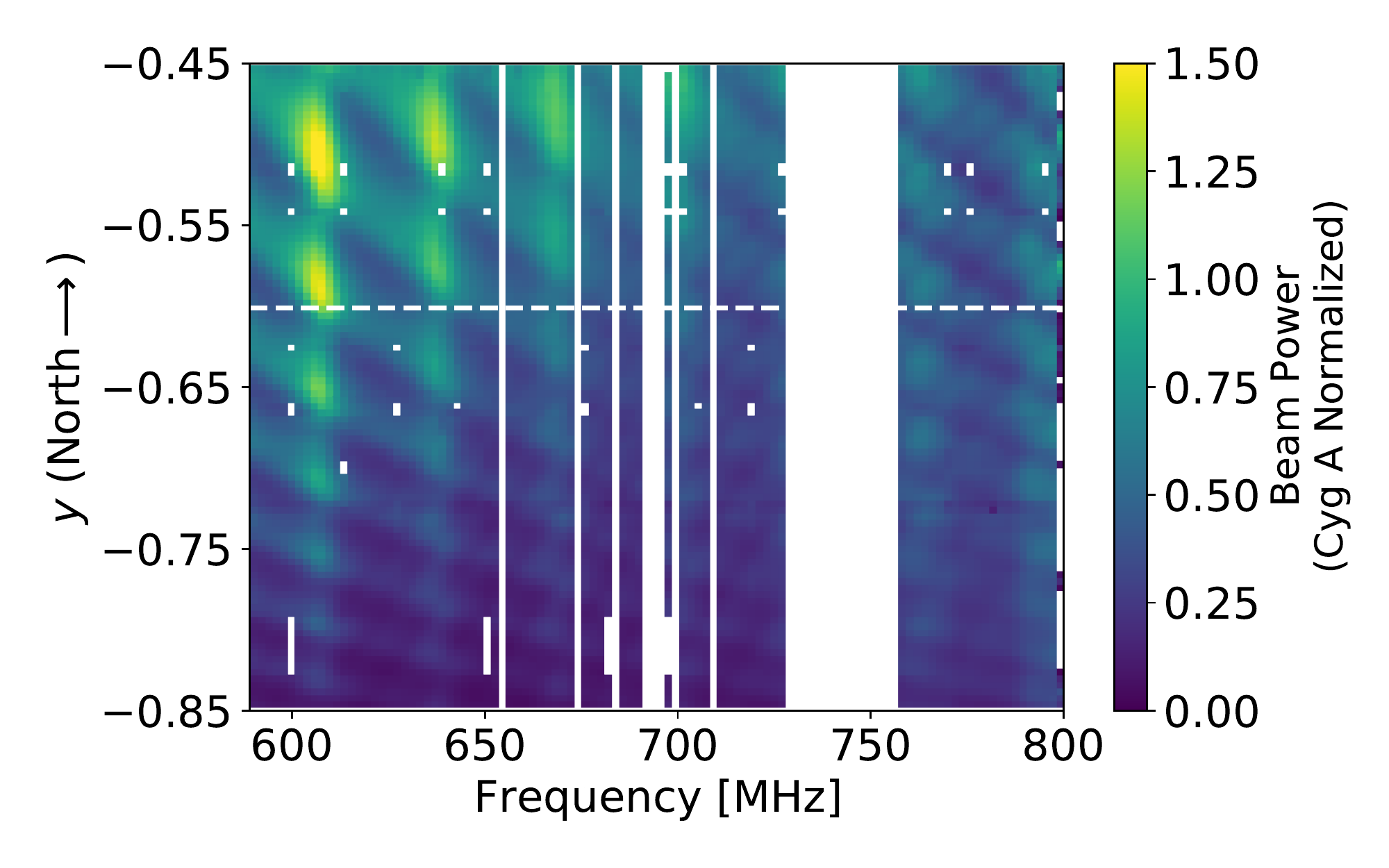}
\hspace{0mm}
\includegraphics[width=0.9\columnwidth]{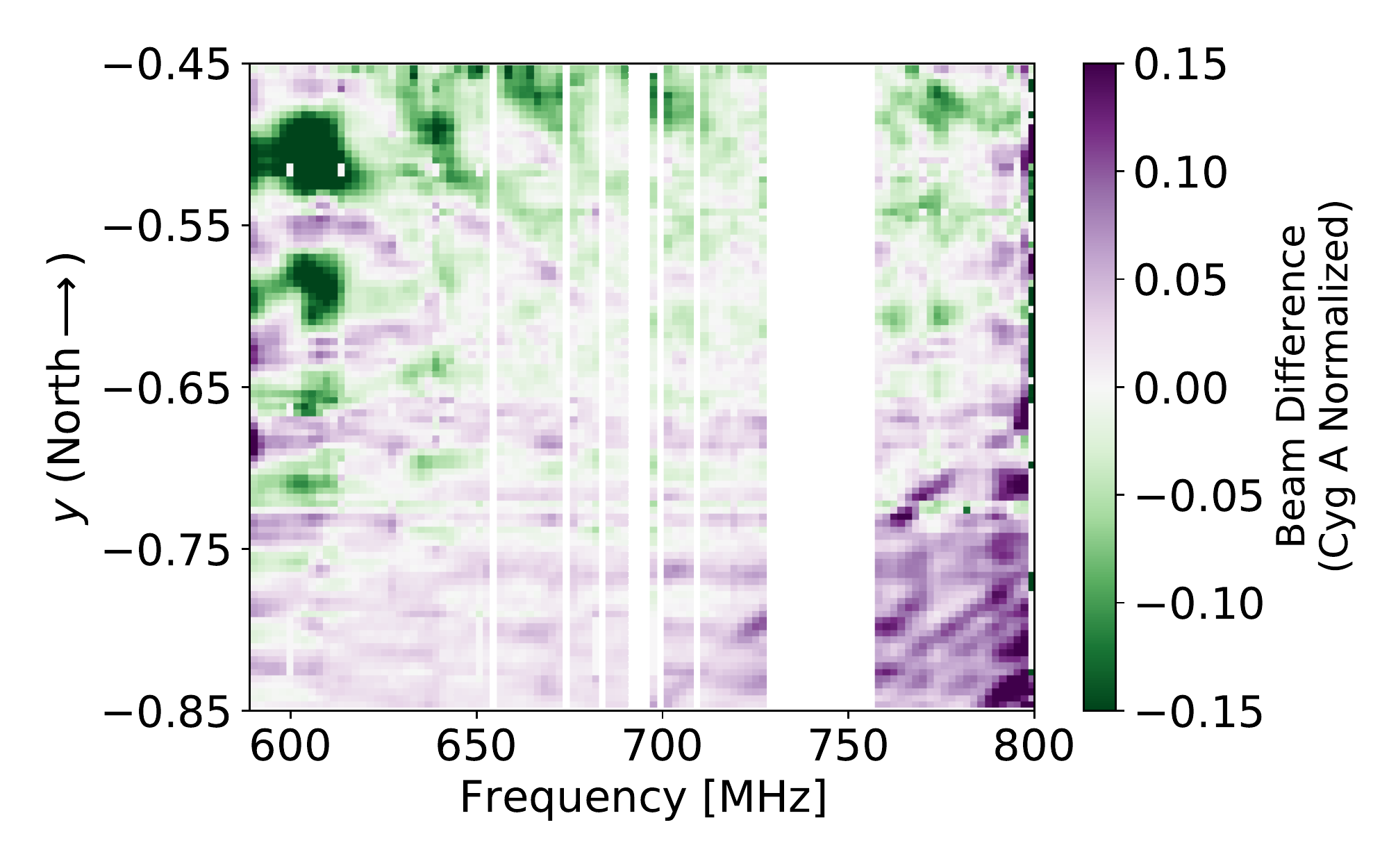}
\caption{The solar measurement of the North-South polarized beam as a function of orthographic coordinate $y$ and frequency (top), and the difference compared the corresponding measurement made by deconvolving a point source model of the sky from a long-baseline CHIME map (bottom). The white dashed line in the top panel marks the position of Sun during the transit shown in Figure~\ref{fig:transit}. For both the North-South and East-West polarized beams (latter not shown), the differences are approximately normally distributed with $\sigma=0.04$ in Cyg~A normalized units.}
\label{fig:decon}
\end{figure}

In Figure~\ref{fig:decon} we plot the North-South polarized beam at $x=0$ as a function of $y$ and frequency, as well as the difference between this beam and the deconvolved point source beam. Currently, the deconvolved point source beam has only been evaluated at frequencies greater than 585~MHz. Again, the level of agreement is typically a few percent of the peak response, though there are some large excursions around 600~MHz. We also find that the solar beam measurement is systematically wider than the deconvolved point source beam by $\sim$5\% in both polarizations.  These differences in width are larger than can be explained by the angular extent of the Sun and efforts to understand the discrepancies between measurements are ongoing. 

\section{Discussion} \label{sec:discuss}

In this section we will discuss the utility of this beam measurement in the context of CHIME science.  As its name suggests, one of CHIME's primary science objectives is to map the distribution of neutral Hydrogen in the low-redshift ($z\sim 1$) Universe. It is difficult to generalize the beam knowledge requirements to meet this objective, since any statement on beam accuracy depends on both the details of the analysis and the form of the beam errors.  To give one example, the analysis presented in \cite{shaw_coaxing_2015} using $m$-mode formalism and a Karhunen–Loève (KL) transform for foreground removal concluded that the beam width had to be known to an accuracy of $10^{-3}$ in order to avoid foreground contamination of the 21~cm power spectrum. For comparison, we found in Section~\ref{sec:compare} that the East-West width of the solar beam measurements were discrepant with other measurements at the level of few percent, or about one and a half orders of magnitude worse than the desired accuracy for this particular analysis. 

However, there are reasons for optimism. First, the modeled beam width errors in \cite{shaw_coaxing_2015} were drawn from a random distribution, resulting in power leakage from low-$k$, foreground-dominated modes to higher $k$ signal modes. In contrast, the errors in this measurement are highly correlated in frequency, affecting the high-$k$ modes to a lesser degree. Second, the use of alternative foreground removal techniques, instead of or in conjunction with the KL transform, may relax requirements on instrumental knowledge. The degree to which both of these factors impact the ultimate analysis is a subject of ongoing investigation.

Nevertheless, these solar beam measurements bring us significantly closer to achieving CHIME's cosmology goals and, in combination with other beam measurements, have enabled the first 21~cm detection with CHIME \citep{stacking_2022}. These measurements can also be used to aid in interpolating between the comparatively sparse point source beam measurements that are outside the solar declination range, bringing us closer to a full $2\pi$ description of the beam. 

As shown in Figure~\ref{fig:allmeas}, the solar measurement is the only data set to provide significant information about the 2D spatial structure of the primary beam.  While the structure appears rather complicated, in fact it is highly separable in $x$ and $y$ and its variance is dominated by a few separable spatial modes at each frequency.  This relative simplicity is being used to guide us in how to interpolate between point source measurements over the rest of the sky.  One approach we are currently exploring is to derive a set of basis functions in $x$ from the solar data, which are then fit to the point source data near meridian, at each $y$ and frequency.  These efforts show promising early results.

In addition to such data-driven analyses, the solar data can also be used to improve our physical understanding of the telescope. As mentioned in Section~\ref{sec:results}, the complicated spatial structure of the beam is qualitatively expected due to multi-path and receiver cross-talk effects within the telescope.  We have been developing a “coupling model” similar to that described by \cite{kern_mitigating_2019} that may be fit to our full set of beam measurements.  The solar data play a key role in constraining the off-meridian properties of this model, which account for 10-20\% of the CHIME’s total primary beam solid angle.

These solar beam measurements are also applicable to CHIME observations of fast radio bursts and pulsars, which have considerably less stringent requirements on beam uncertainty compared to the cosmological measurement. The beam models used in \cite{amiri_first_2021} and \cite{good_first_2021} were each informed in part by these data. Also, in combination with a holographic measurement of Tau A, these data were used to calibrate the spectral flux density of the flare observed from Galactic magnetar SGR 1935+2154 reported in \cite{andersen_bright_2020}.

\section{Conclusions}

We have presented a measurement of the CHIME primary beam pattern using the Sun as a calibration source. This data set is an integral part of the broader CHIME beam measurement program, complementing more conventional beam measurements using astrophysical point sources.  This data set provides a direct measurement of the CHIME beam at low declination and is indeed the \textit{only} measurement over much of the sky.  Where overlap exists with other beam measurements, comparisons can be made to constrain the systematics of different measurement techniques.  The rich spatial structure revealed by this measurement has enormous qualitative and quantitative value for understanding the source of beam features and may enable extrapolating beyond the current measurement.  Efforts to combine beam measurements will be the topic of future publications.

The semiannual motion of the Sun enables measuring a large solid angle without the need to move the telescope, which is particularly useful for driftscan instruments. Despite the intrinsic variability of the Sun, the uncertainty on the resulting beam measurement is estimated to be $<10\%$  where the response is above the confusion limit, which is $\sim40$~dB below the peak beam response. While CHIME was able to take advantage of the solar minimum period for these observations, future instruments wishing to use the Sun for beam measurements may find themselves poorly situated with respect to the 11-year solar cycle.  However, solar beam measurements may still be viable in times of increased solar activity. Whereas this analysis made no attempt to correct for variations in the solar flux, even simple adjustments based on independently available solar monitor data may be sufficient to account for the slowly varying component of solar emission.

\begin{acknowledgments}
We thank the Dominion Radio Astrophysical Observatory, operated by the National Research Council Canada, for leasing the CHIME site and for their gracious hospitality and support. DRAO is situated on the traditional, ancestral, and unceded territory of the Syilx Okanagan people. We are fortunate to live and work on these lands. CHIME is funded by grants from the Canada Foundation for Innovation (CFI) 2012 Leading Edge Fund (Project 31170), the CFI 2015 Innovation Fund (Project 33213), and by contributions from the provinces of British Columbia, Qu\'ebec, and Ontario. Long-term data storage and computational support for analysis is provided by WestGrid\footnote{\url{https://www.westgrid.ca}} and Compute Canada\footnote{\url{https://www.computecanada.ca}}. 

Additional support was provided by the University  of British Columbia, McGill University, and the University of Toronto. CHIME also benefits from NSERC Discovery Grants to several researchers, funding from the Canadian Institute for Advanced Research (CIFAR), and from the Dunlap Institute for Astronomy and Astrophysics at the University of Toronto, which is funded through an endowment established by the David Dunlap family. This material is partly based on work supported by the  NSF through  grants (2008031)  (2006911) and  (2006548) and by the Perimeter Institute for Theoretical Physics, which in turn is supported by the Government of Canada through Industry Canada and by the Province of Ontario through the Ministry of Research and Innovation.

\end{acknowledgments}

%

\vspace{5mm}
\facilities{CHIME \citep{the_chime_collaboration_overview_2022},
    Learmonth Solar Observatory \citep{SWS},
    John A. Galt Telescope \citep{locke_1420_1965}}


\software{   NumPy \citep{NumPy},
   SciPy \citep{SciPy},
   HDF5 \citep{HDF5},
   h5py \citep{h5py},
   Matplotlib \citep{Matplotlib},
   Skyfield \citep{Skyfield},
   caput \citep{caput},
   ch\_pipeline \citep{ch_pipeline},
   draco \citep{draco},
   kotekan \citep{kotekan}}


\bibliography{Solar}{}
\bibliographystyle{aasjournal}



\end{document}